\documentstyle[epsfig]{l-aa}
\newcommand{\ltsima} {$\; \buildrel < \over \sim \;$}
\newcommand{\gtsima} {$\; \buildrel > \over \sim \;$}
\newcommand{\lta} {\lower.5ex\hbox{\ltsima}}
\newcommand{\gta} {\lower.5ex\hbox{\gtsima}}

\def\ltsima{$\; \buildrel < \over \sim \;$}
\def\simlt{\lower.5ex\hbox{\ltsima}}
\def\gtsima{$\; \buildrel > \over \sim \;$}
\def\simgt{\lower.5ex\hbox{\gtsima}}

\def \rsun {\ifmmode$R$_{\odot}\else R$_{\odot}$\fi}
\def \nh {N${\rm _H}$}
\def \hcm {\hbox {\ifmmode $ H atoms cm$^{-2}\else H atoms cm$^{-2}$\fi}}
\def\approxgt{\mathrel{\hbox{\rlap{\lower.55ex \hbox {$\sim$}}
        \kern-.3em \raise.4ex \hbox{$>$}}}}
\def\approxlt{\mathrel{\hbox{\rlap{\lower.55ex \hbox {$\sim$}}
        \kern-.3em \raise.4ex \hbox{$<$}}}}
\newcommand {\rosat} {{ROSAT}}

\newcommand {\asca} {{\it ASCA}}
\newcommand {\sax} {{\it BeppoSAX}}

\begin{document}
\thesaurus{ }

%
   \title{Reprocessing and Variable Cold Absorption in  the Broad-Line Radio 
Galaxy 3C~390.3}

   \author{P. Grandi\inst{1} \and
   M. Guainazzi\inst{2} \and  F. Haardt\inst{3}
\and L. Maraschi\inst{4} \and
E. Massaro\inst{5} \and G. Matt\inst{6} \and
L. Piro\inst{1} \and C. M. Urry\inst{7}}

\institute{
{Istituto di Astrofisica Spaziale, C.N.R., Area Ricerca di Roma Tor Vergata
I-00133 Roma, Italy}
\and
{Astrophysics Division, Space Science Department of ESA, ESTEC, Postbus
 299, 2200 AG Noordwijk, The Netherlands}
\and
{Dipartimento di Fisica, Universit\`a di Milano, Via Celoria, I-20133 Milano,
Italy}
\and{Osservatorio Astronomico di Brera, Via Brera 28, I-20121 Milano, Italy}
\and{Istituto Astronomico, Universita' di Roma, Unita' GIFCO-CNR Roma I, Via
Lancisi 29 Roma, Italy}
\and
{Dipartimento di Fisica ``E.Amaldi'', Universit\`a degli Studi ``Roma 3'',
Via della Vasca Navale 84, I-00146 Roma, Italy}
\and{Space Telescope Science Institute, 3700 San Martin Drive, Baltimore,
MD 21218, USA}}

   \offprints{P. Grandi}

\date{Received  1998 May 14; accepted 1998 October 26}
\maketitle

\markboth{P.Grandi et al.}{\sax~observation }

\begin{abstract}

A \sax~observation of the Broad Line Radio--Galaxy 3C~390.3 is reported. For
the first time,  both the K$_{\alpha}$ iron line and a strong reflection
hump, produced by the illumination of the primary X--ray emission on cold
matter, are detected in this source. The 0.1--100 keV  continuum is modeled by
an absorbed hard power law ($\Gamma\sim 1.8$) reflected at high energies by
material with a fairly large covering factor ($\Omega/2\pi \simeq 1$). The iron
line is centered at $\simeq 6.4$ keV (rest frame), is intrinsically narrow
($\sigma=73^{+207}_{-73}$ eV), and has an equivalent width of $\simeq 140$ eV. 

We discuss the results in the context of current models for AGNs and suggest
that the primary  X--ray power law continuum is probably produced by a hot
inner flow, while the reprocessed radiation comes from an outer cold thin disk,
and/or from a thick torus at even larger radii. Further observations with \sax~
could distinguish between the latter two cases. Beamed radiation associated to
the radio jet is unlikely to  contribute significantly to the X--ray emission. 

Finally, an historical study of the column density \nh, also reported here,
shows that the absorption along the line of sight changes in time. The \nh~ time
variability, which is not correlated with that of the primary continuum, seems to imply
variations of the geometry of the absorber rather than variations in the
ionization state of the gas. 

\end{abstract}

\section{Introduction}

In recent years, statistical studies of complete samples of extragalactic radio
sources have suggested a unified scheme for radio--loud AGN. According to this
picture, radio galaxies, radio--loud quasars, and blazars are the same physical
objects seen at decreasing angles with respect to the jet axis (Antonucci 1993;
Urry $\&$ Padovani, 1995). A similar scenario has been proposed for
radio--quiet AGN (i.e., Seyfert 1 and Seyfert 2 galaxies; Antonucci \& Miller
1985; Miller \& Goodrich 1990). The common motivation for such unified schemes
is that the radiation field is almost certainly {\it anisotropic}, which
automatically implies orientation--dependent observational properties. Some
degree of anisotropy could be caused by an opaque circumnuclear torus, which
for some lines of sight may prevent direct view of the active nucleus and of
the broad emission line region. In radio--loud AGN, additional anisotropy is
likely due to relativistic beaming of the continuum produced in the jet. 

Broadly speaking, a radio--loud source should show a featureless (or almost
featureless) spectrum when observed face--on, a Seyfert 1--like spectrum at
intermediate angles and a Seyfert 2--like spectrum when seen edge--on. Broad
line Radio Galaxies (BLRG) and  Narrow Line Radio Galaxies (NLRG) are therefore
considered the radio--loud counterpart of Seyfert 1 and Seyfert 2 galaxies,
respectively. However, it is still unclear whether 
 the accretion processes are the same in radio--loud AGN and in Seyfert galaxies.
Indeed, Rees et al. (1982) suggested that in radio galaxies the accreting gas
flow is not in a cold, geometrically thin, disk configuration, but it rather forms a
hot, geometrically thick, ion--supported torus, characterized by low radiative
efficiency. The Advection Dominated Accretion Flow (ADAF) models, more recently
proposed by other authors (see Narayan, Mahadevan \& Quataert 1998, hereinafter
NMQ, for a recent review), follow similar lines of thought. Shapiro, Lightman
\& Eardley (SLE, 1976) found a solution to the accretion problem that, in many
respects, resembles the later ion--supported torus, and suggested 
its relevance for the black hole candidate Cygnus X--1.
In the SLE solution, however, the energy produced in the flow by
viscosity is locally radiated and advection is implicitly assumed to be
negligible. A hot ion torus, surrounded by a geometrically thin cold accretion disk
irradiated by the hard radiation produced by the torus,
 was proposed by Chen and Halpern (1989) in the context of BLRG showing double
peaked emission lines, notably Arp 102B and 3C~390.3. 
 
From the observational point of view the situation  is still confused. 
{\it Ginga} data of radio--loud objects
showed uncertain detection of the iron line and/or the reflection component
(Nandra $\&$ Pounds 1994). {\it ASCA} observations at better energy resolution showed
the presence of the iron line in several radio--loud AGNs, but did not
constrain the reflection hump (Eracleous et al. 1997, Allen $\&$ Fabian 1992,
Grandi et al. 1997a, Yamashita $\&$ Inoue 1996). It is then  unclear whether
a Seyfert--like nucleus is present in radio galaxies. The wide energy band
($\sim 0.1--100$ keV) covered by the instruments on board \sax~ is
particularly appropriate to address the problem. For this reason, a \sax~Core
Program has been dedicated to the spectral study of bright radio galaxies
(F$_{\it 2-10 keV}>10^{-11}$ erg cm$^{-2}$ sec$^{-1}$). Here we present the
observation of 3C~390.3, the first source observed. 

3C~390.3 is a well known Broad--Line Radio Galaxy (z=0.057) with an
FRII morphology and a core showing superluminal motion (Alef et al. 1996). Its
spectrum is characterized by double--peaked emission lines in the optical and UV
bands (Eracleous and Halpern 1994, Zheng 1996, Wamsteker et al. 1997). The UV
bump, typically observed in most Seyfert galaxies, is weak or even absent
(Wamstecker et al. 1997). The Einstein--IPC data revealed the presence of a
strong intrinsic absorption (Kruper et al. 1990). Ghosh and Soundararajaperumal
(1991) claimed the presence of a soft excess in the EXOSAT data, but their
results have not been confirmed by the later \rosat~and {\it ASCA} observations
(Walter et al. 1994, Leighly et al. 1997). At higher energies, analysis of
{\it Ginga} data produced ambiguous results. Inda et al. (1994) resolved the iron
line at 6.4 keV but not the reflection component. Nandra $\&$ Pounds (1994)
could give only an upper limit on the equivalent width of the emission line,
but revealed a weak reflection component of small covering factor
($\Omega/2\pi\sim 0.4$). {\it ASCA} confirmed the presence of the iron K emission
line (Eracleous et al. 1996, Leighly et al. 1997), but could not constrain the
reflection hump, most probably because of the limited energy range. 3C~390.3
has also been detected by OSSE in the soft $\gamma-$ray domain, above 50 keV
(Dermer $\&$ Geherels 1995) . 
 
\section{Data Reduction}

3C~390.3 was observed on Jan. 9--10, 1997, with the Narrow Field Instruments
(NFI) of the \sax~ satellite (Parmar et al. 1997, Boella et al. 1997, Manzo et
al. 1997 and Frontera et al. 1997). The data reduction followed the standard
procedure. LECS and MECS cleaned photon lists and PDS background--subtracted
products (spectrum and light curve) were obtained using {\tt SAXDAS 1.3.0},
included in the {\tt Ftools} package. 

The LECS and the MECS (3 units) spectra were accumulated over circular regions
of 4$\prime$ radius and the background spectra were extracted from blank field
observations, using extraction regions on the detector equal to those of the
source. The observed background--subtracted count rates are 0.18$\pm0.03$,
0.37$\pm0.02$ and 0.49$\pm0.032$ with a net exposure time of 35 ksec, 100 ksec
and 46 Ksec in the LECS (0.12--10 keV),  MECS (1.5--10 keV) and PDS (13--100
keV), respectively. 

The  grouping files available at the \sax~ Scientific Data Center
(http://www.sdc.asi.it/software/) were used to re--bin the data
with the ftool GRPPHA. The data were
rebinned so as to allow the use of the $\chi^2$ statistic and, at the same
time, to sample the spectral resolution of the instruments ($\Delta
E/E=8\times(E/6)^{-0.5} FWHM \%$ for the LECS and the MECS; $\Delta
E/E=0.15\times(E/60)^{-0.5} FWHM \%$ for the PDS). Publicly available matrices
(September 1997 release) were used for all the instruments. 

The LECS, MECS and PDS light curves were inspected to search for possible time
variability, using the {\tt XRONOS} package. A standard $\chi^2$ test was
applied to the average count rate in each light curve. We could not detect any
flux variation in the whole 0.1--100 keV range during our observation. The
$\chi^2$ probability that the source was not constant is smaller than $10^{-3}$
in each instrument, independently of the temporal bin size used. This result
confirms observations done with older X--ray missions, which never detected
short time variability on time scales $\lta 1$ day (Shafer, Ward \& Barr
1985; Ghosh \& Soundararajaperumal 1991; Inda et al. 1994; 
Leighly et al. 1997). 

\section{Spectral Analysis}

The 3C~390.3 data were initially divided by the 3C 273 data collected during
the \sax~ Science Verification Phase (SVP). This method allows a
quick qualitative look at the spectrum without the need for complex models. The
SVP spectrum of 3C 273 was chosen because it is well represented  by a simple power
law in the 1.5--200 keV. Only a very small deviation occurs at $\sim 5.4 $ keV
(observer frame) due to the presence of a weak iron line (EW$\sim30$ eV). The
LECS data were not considered because of the spectral complexity of 3C 273
below 1 keV (Grandi et al. 1997b). The spectral ratio immediately evidenced the
presence of a line at $\sim 6.0$ keV (as expected from neutral iron, taking
into account the redshift of the source), and a clear excess above 10 keV.

Because of the complexity of the 3C~390.3 spectrum, we  decided to perform the
spectral analysis in two steps, studying the continuum and the emission line
separately. Special care was also taken in determining the absorbing column
density along the line of sight. 

\subsection{\it Continuum: primary emission and reflection component}

We fitted the \sax~ data in XSPEC between 0.12 and 100 keV excluding the points
in the energy interval 5.5--7.0 keV, where the emission line contributes
significantly, aiming for the best possible determination of the continuum. The
relative flux normalization between the LECS and MECS was left free to vary,
whereas a miscalibration of 15$\%$ between the MECS and the PDS was assumed as
indicated by NFI intercalibration analysis based on observations of 3C 273
(Grandi et al. in preparation). 

We first tested a single power law model (PL), with low energy absorption. The
best fit parameters are reported in Table 1, with 90$\%$ confidence limits.
This model is unacceptable as it gives a large value of $\chi^2$,
corresponding to a chance probability of about 0.004. A broken power law model
(BKP) gives a better fit (see Table 1), but still not good enough in terms of
the $\chi^2$ statistic (chance probability of 0.054). 

We therefore tested a more complex spectral model including, in addition to a
 power law, a reflection component (PEXRAV in XSPEC; see Magdziarz
$\&$ Zdziarski 1995). The PEXRAV model assumes that a cold reflector is
irradiated by a primary isotropic X--ray source. The input X--ray spectrum can
be modeled by a simple power law or by a power--law with a high energy
exponential cut--off. The general functional form for this model is: 

$$N(E)=A*[E^{-\Gamma}*e^{-E/E_{c}}+R*f_{ref}(E,i,\Gamma,E_{c})],$$ 

\noindent
where A is a global normalization factor at 1 keV (photons cm$^{-2}$ sec$^{-1}$
keV$^{-1}$), $\Gamma$ the photon index, and $E_{c}$ the cutoff e--folding
energy. Setting $E_{c}$=0, the exponential term is excluded from the fitting
formula, and a simple power law is taken as primary continuum. The reflection
component $f_{ref}$ is computed assuming a plane parallel semi--infinite medium
irradiated by a point--like or optically thin X--ray source, and    is a
function of the angle $i$ between the line of sight and the normal to the slab.
We fixed $i$ to 26$^o$ as deduced from the UV and optical emission line
measures (Eracleous $\&$ Halpern 1994 and Wamsteker et al. 1997), and from the
radio jet superluminal motion (Eracleous et al. 1996). Finally, R is an additional
 scaling factor introduced to take into account  roughly
 the solid angle subtended by the cold reflecting
material to the X--ray source located above it, when different from 2$\pi$. 
R=1 corresponds to the scale free geometry  assumed in the computation of $f_{ref}$.
In principle, a change of the subtended
solid angle would imply a change in the angular distribution of impinging
photons, which in turn would change not just the {\it normalization} of the
reflected spectrum, but also its {\it shape}. However, such an effect is  small in
comparison to the quality of available data,
 and the geometry is usually assumed to be simply described by the parameter
R = $\Omega/2\pi$. 

The MECS and PDS data are better reproduced by the model including reflection
than by the single (PL) or broken power law (BKP) models. We initially assumed
a  simple  power law input spectrum (PL+REF). The reduced $\chi^2$ decreases
from 1.17 to the well acceptable value of 1.09 (chance probability of 0.17).
The reflection component is unambiguously detected at 99$\%$ confidence level,
as evident from the contour plots shown in Figure 1. We then included a high
energy cutoff (CPL+REF), allowing $E_{c}$ to vary. As expected, when the  input
X--ray spectrum is a power law with cut--off (E$_{c}\sim 400$ keV), the deduced
value of R  (R=1.2) is somewhat larger, but the increase is not statistically
significant. In fact, in order to reproduce the observed Compton hump, the
reflection component has to be a bit boosted to compensate for the reduction of
primary  photons at very high energies, which preferentially emerge,  once
down--scattered in the cold layers, around 20 keV.  In any case, the
improvement in $\chi^2$ is not significant ($\Delta\chi^2=3$), and
therefore the inclusion of a cut--off is not statistically required by the
data. We note, however, that non--simultaneous {\it ASCA}, {\it Ginga}, and OSSE
observations (Wozniak et al. 1998) suggest the presence of a break in the high
energy spectrum of 3C~390.3. Also, the spectral index variations measured by
{\it ASCA} (Leighly et al. 1997) indicate a pivot point at $\sim 400$ keV. 

In any case, the amount of reflection required by the \sax~data is larger than
that previously measured by {\it Ginga} (R$\sim 0.3-0.4$, (Nandra \& Pounds 1994,
Wozniak et al. 1998), possibly suggesting variations of the relative strength
of the reflection hump with respect to the primary continuum. 

\begin{figure}[t]
\label{fig1}
\epsfig{figure=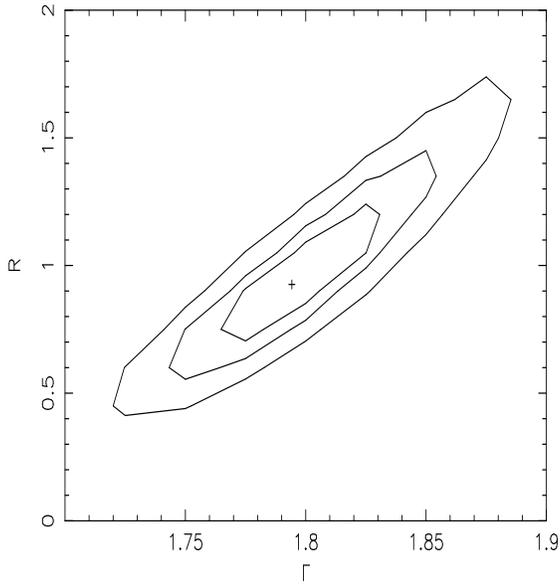,height=8.5cm,width=8.0cm,angle=-90}
\caption{Confidence contours for the photon index ($\Gamma$) and  the 
amount of reflection (R) when the continuum is fitted with a power law 
reflected by cold matter (model PL+REF in Table 1).}
\end{figure}

\begin{table*}
\begin{center}
\caption[] {Fits to \sax~ continuum$^a$ of 3C~390.3}. 
\begin{tabular}{lccccccc}
\noalign {\hrule}
&&&&&&&\\
   &A$^b$ & \nh (10$^{20}$ cm$^{-2}$) & $\Gamma$
     & $\Gamma_{hard}^c$ & E$_{break}^d$ & R & $\chi^2$ (d.o.f)         \\
&&&&&&&\\
&&&&&&&\\    
PL  &5.1$^{+0.2}_{-0.1}$ &10$^{+2}_{-1}$ &1.68$\pm0.02$ & - & - & 0 & 
245 (190)\\
&&&&&&&\\
BKP &5.4$\pm0.2$ &12$^{+1}_{-2}$&1.72$^{+0.04}_{-0.03}$ 
& 1.49$\pm0.07$ & 6.7$\pm 1.4$ & 0 & 220(188)\\
&&&&&&&\\
PL+REF &5.6$\pm.3$ &13$\pm2$ &1.80$\pm0.05$ & - & - 
& 0.9$^{+0.5}_{-0.3}$& 207.6(189)\\
&&&&&&&\\
CPL +REF & 5.6$^{+0.4}_{-0.3}$  & 13$\pm2$ & 1.80$^{+0.05}_{-0.04}$  & 
- & 380$^{+\infty}_{-257}$ & 1.2$^{+0.4}_{-0.3}$ & 204.5(188)\\
&&&&&&&\\\hline
\multicolumn{8}{l}{$^a$ Data in the 5.5--7 keV energy band
are not included in the fits.}\\
\multicolumn{8}{l}{$^b$ Normalization factor at 1 keV ($\times$ 10$^{-3}$
photons cm$^{-2}$ sec$^{-1}$ keV$^{-1}$).}\\
\multicolumn{8}{l}{$^c$ High--energy power law for model BKP.}\\
\multicolumn{8}{l}{$^d$ Power--law break in model BKP, e--folding cut--off 
energy in model CPL+REF.}
\end{tabular}
\end{center}
\end{table*}
\noindent

\subsection{\it Column Density}

The low energy absorption requires a column density larger than the
Galactic value, N$_{\rm H}^{Gal}=4.2\times10^{20}$ cm$^{-2}$ (Stark et al.
1992), for every model of the continuum we tested. Comparing our results with
ROSAT data (Wamsteker et al. 1997), we noted that our determination of the
column density is slightly larger. We then collected all the available
information from the literature in order to investigate possible temporal
variations of \nh. We re--analyzed two old EXOSAT observations performed in
1985 Feb. 3 and 1996 Mar. 17--18, respectively, as they provide a
quasi--simultaneous coverage of the 0.01--10 keV band, and have a good
signal--to--noise in the ME spectra (quality flag $ \ge 3$). The ME (1.8--10.0
keV) and LE (0.01--2 keV) data were simultaneously fitted with a simple power
law with cold photoelectric absorption, which turned out to provide an adequate
fit to the data in both the observations. The best fit values are reported in
Table 2. 
\begin{figure}[ht]
\label{fig2}
\epsfig{figure=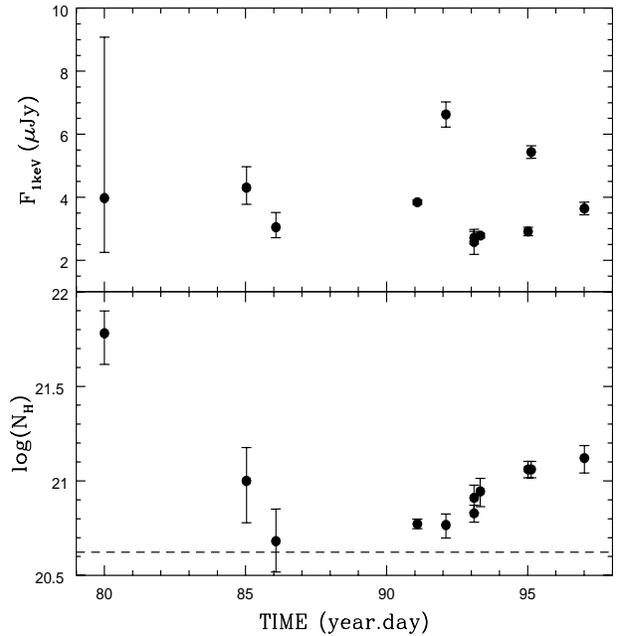, height=9.0cm, width=8.5cm }
\caption{ ({\it upper panel}) -- Historical X--ray light curve of 3C~390.3.
({\it lower panel}) -- The column density is plotted as a function of 
time. 
The \nh~ trend was decreasing before 1987 and increasing 
after 1991. The dashed  line corresponds to the Galactic column density}   
\end{figure}
No soft  X--ray excess is required by the EXOSAT data. The LE data points lie
on the extrapolation of the power--law continuum, which is basically determined
by the ME statistics. This result is in disagreement with Ghosh and
Soundararajaperumal (1991), who claimed that a soft excess is present in this
source. 

\begin{table*}
\begin{center}
\caption[]{Best--fit parameters when a simple absorber power--law model is
applied to the Feb. 3, 1985, and Mar 17--18, 1986, EXOSAT observations of
3C~390.3}
\begin{tabular}{lcccc} \hline 
\noalign {\hrule}
Year & Instrument & \nh & $\Gamma$ & $\chi^2$ (d.o.f.) \\ 
     &            &        
($10^{20}$~cm$^{-2}$) & &\\
\hline
&&&&\\
1985 &  ME+ AL + 3LEX & $10^{+5}_{-3}$ & $1.61 \pm 0.10$ & 50(43) \\
&&&&\\
1986 &  ME+ AL + 3LEX & $4.6^{+2.6}_{-1.5}$ & $1.56^{+0.10}_{-0.08}$& 25(23) \\ 
&&&&\\
\hline 
\noalign {\hrule}
\end{tabular}
\label{tabexosat}
\end{center}
\end{table*}

When the historical column density is plotted as a function of the date of
observation, several interesting variations are evident, clearly
indicating that temporal modifications of the cold material along the line of
sight occured, with an estimated time scale of a few years (see Figure 2). In
particular, the value of \nh~seems to be steadily increasing since 1992,  but
was always found much lower than the first Einstein detection (Kruper et al.
1990). 

We did not find any correlation between \nh~ and the intensity at 1 keV (see
Figure 2 and Table 3). This fact suggests that changes in the geometry of the
absorber, rather than variations of its ionization state (as expected in the
case of a warm absorber responding to an ionizing continuum), are probably
responsible for the observed long term variability of \nh. 
Similar long--term absorption variability has been also  observed in 
NGC4151 (Yaqoob et al. 1993).
\begin{figure}[h]
\label{fig3}
\epsfig{figure=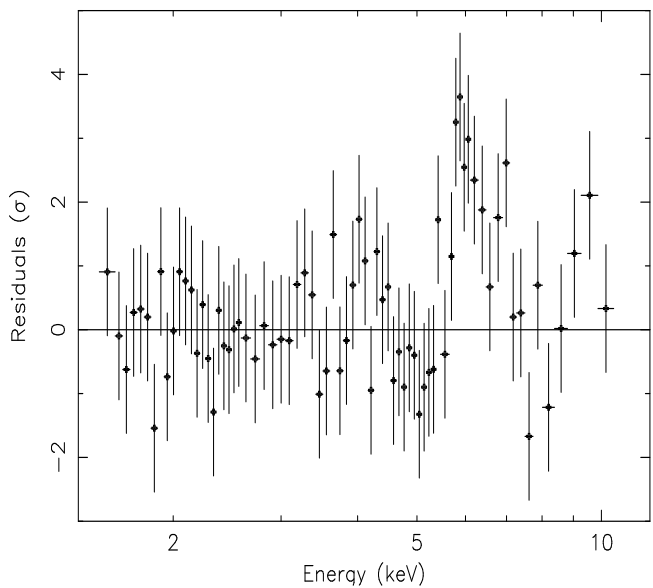,height=8.5cm,width=9.5cm,angle=-90}
\caption{MECS residuals ($\sigma$) to the data, when the
continuum of emission s fitted with a power law reflected by cold material.
The error bar corresponds to one sigma}
\end{figure}
\subsection {\it Emission Line}

An excess in emission with respect to the power law plus reflection continuum
is clearly present in the 5.5--7.0 keV energy interval (see Figure 3). 
The statistical significance of this excess, estimated by the quadratic sum of the
deviations in each bin, is at the level of $\simeq 7 \sigma$. The excess is
most likely produced by an Fe fluorescent line, for which we could estimate,
using a gaussian profile, a flux of 3.63 (-1.45,+0.78) $\times 10^{-5}$ ph cm$^{-2}$
sec$^{-1}$ (at the 90$\%$ confidence level for one interesting parameter),
corresponding to an equivalent width of 136 (-36,+40) eV. The line centroid
lies at 6.39 (-0.09,+0.10) keV (rest frame), compatible with K$\alpha$ line
emission from neutral iron located at the source redshift. The line is not
resolved: the intrinsic width can be estimated from the fits as 73 (-73,+207)
eV, indicative of a narrow feature. The line flux and the intrinsic width
confidence contours are shown in Figure 4.
\begin{figure}[th]
\label{fig4}
\epsfig{figure=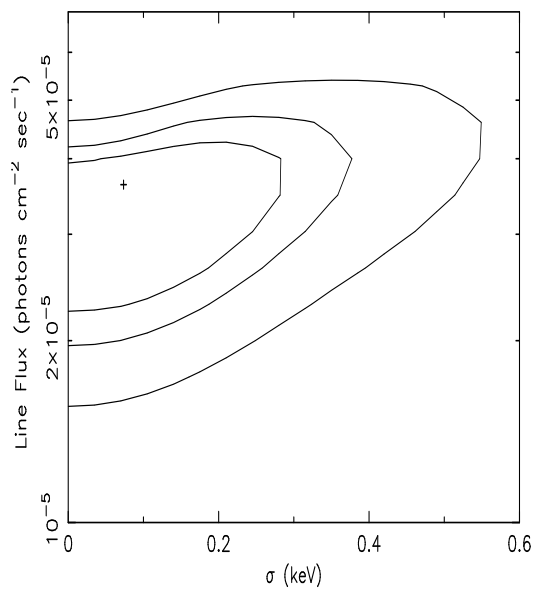,height=8.5cm,width=7.5cm,angle=-90}
\caption{ Confidence contours for the line flux
and the intrinsic width when the feature is fitted with a
gaussian profile.}
\end{figure}
These line best fit values are consistent
with those of previous {\it ASCA} observations (Eracleous et al. 1996, Leighly et al.
1997): in fact  the line intensity derived from either {\it Ginga}, {\it ASCA} or \sax~ is
consistent with being constant within the rather large errors (cf. Wozniak et
al. 1998). 

In Figure 5 the total 0.1-100 keV photon spectrum is shown
with the residuals to a power--law--plus--reflection model when the gaussian line is
added to the fit.
\begin{figure}
\label{fig5}
\epsfig{figure=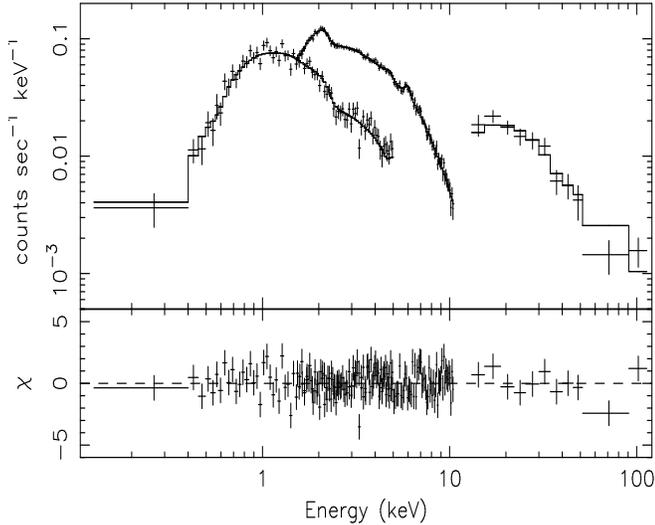,height=8.5cm,width=9.5cm,angle=-90}
\caption{Photon spectrum ({\it upper panel}) and
residuals ({\it lower panel})
when a power law plus reflection and a gaussian line are fitted to
to 0.12-100 keV \sax~data}
\end{figure}

\section{Discussion}

The detection of the iron line and of the reflection component in the
\sax~observation of 3C~390.3 indicates that beamed non--thermal radiation does
not contribute significantly to the X--ray continuum. This is probably true
independent of the brightness of the X--ray source, because a strong iron
line ($EW\sim 130$) was also detected by {\it ASCA} in 1995 when the source flux was
about 1.5 times larger than the \sax~ value (Leighly et al. 1997). 
\begin{table*}
\begin{center}
\caption[] {Absorbing Column Density History}
\begin{tabular}{lcccl}
\noalign {\hrule}
Satellite/Instrument & Date of obs. & \nh (cm$^{-2}$)& 
Flux$_{\it 1 keV}$ & References\\

&  Year.day & ($\times10^{20}$)& $\mu$Jy &\\
&&&&\\\hline
EINSTEIN/IPC & 80.001& 60$\pm19$ & 4.0$^{+5.1}_{-1.7}$ & Kruper 
et al. 1990\\
EXOSAT/LE+ME & 85.033& 10$^{+5}_{-3}$ & 4.3$^{+0.7}_{-0.5}$&this paper\\
EXOSAT/LE+ME & 86.076 &  $4.6^{+2.6}_{-1.5}$& $3.1^{+0.5}_{-0.3}$
& this paper\\
ROSAT/PSPC & 91.089 & 5.9$\pm0.4$&3.8$\pm0.1$& Wamsteker et al. (1997)\\
ROSAT/PSPC  & 92.101 & 5.8$^{+0.8}_{-0.7}$& 6.6$\pm0.4$ &  this paper\\
ROSAT/PSPC & 93.101 & 6.7$\pm0.7$& 2.6$\pm0.4$ &  Wamsteker et al. (1997)\\
ROSAT/PSPC & 93.102 & 8.1$\pm1.4$& 2.7$\pm0.2$ &  Wamsteker et al. (1997)\\
{\it ASCA}/SIS   & 93.320 & 8.8$\pm1.5$&  2.8$\pm0.1$ &  Eracleous et al. (1996)\\
{\it ASCA}/SIS   & 95.015 & 11$\pm1$&  2.9$\pm0.1$ &  Leighly et al. (1997)\\
{\it ASCA}/SIS   & 95.125 & 11$\pm1$&  5.4$\pm0.2$ &  Leighly et al. (1997)\\
BeppoSAX/LE+ME+PDS &97.009& 13$\pm2$&3.6$\pm0.2$& this paper\\
&&&&\\
\noalign {\hrule}
\end{tabular}
\end{center}
\end{table*}
\noindent

In 1995, 3C~390.3 was the object of a multifrequency campaign which included
IUE, ROSAT and {\it ASCA} observations (Leighly et al. 1997, O'Brien et al. 1998).
The UV and X--ray light curves, covering a period of about 8 months with a regular
3 day sampling, showed
similar forms and variability amplitudes.
 As pointed out by
O'Brien \& Leighly (1997), if the UV were a direct extension of the X--ray
emission, the two light curves should show different
variability amplitudes, because the {\it ASCA} spectral slopes from two
observations during the monitoring differed by $\Delta\alpha=0.1$. 

It is therefore likely that (at least part
of) the UV is due to reprocessing of X--rays.
Indeed, an excess of UV emission above the X--ray power law extrapolation (the
blue bump) was noted by Walter et al. (1994) using simultaneous ROSAT--IUE
observations performed during the ROSAT all--sky survey. However, the blue bump
component, if present, is weak, as also indicated by the historical
compilation of non--simultaneous ultraviolet and X--ray data of Wamsteker et
al. (1997). The lack of a soft X--ray excess attested by several satellites
(Walter et al. 1994, Eracleous et al. 1996, Leighly et al. 1997) and confirmed
by our data (and by the re--analysis of the EXOSAT observations) further
strengthens this conclusion. 

In Figure 6 the radio to $\gamma$--ray energy distribution of the 3C~390.3 is
shown. Data from the literature (Rudnick et al. 1986, Steppe et al. 1988, 
Knapp et al. 1990, Poggioli 1991)
are combined with the simultaneous optical--UV--X-ray data 
collected on 1995 January 14-15 during a 3C~390.3 multifrequency campaign
(Leighly et al. 1997, Dietrich et al. 1998, O'Brien et al. 1988).
The radio points correspond to the core flux only. Optical and UV
measurements collected on 1995 January 14-15 refer 
to the continuum emission dereddened with the extinction curve of Seaton (1979)
assuming Av=0.708. The visual extinction was deduced by the \nh~ column 
density measured by a simultaneous {\it ASCA} 
observation performed on 1995 January 15 (Leighly et al. 1997).
On that occasion 3C~390.3 was in a state of brightness very similar to that 
observed by \sax~later, as can be seen in Figure 6, where
the \asca~flux at 1 keV is plotted together with 
the MECS and PDS data.
For the sake of clarity, the MECS and PDS 
data have been rebinned in order to have a signal to noise ratio 
of about 50 and 10 per each bin, respectively. 

\begin{figure}
\label{fig6}
\epsfig{figure=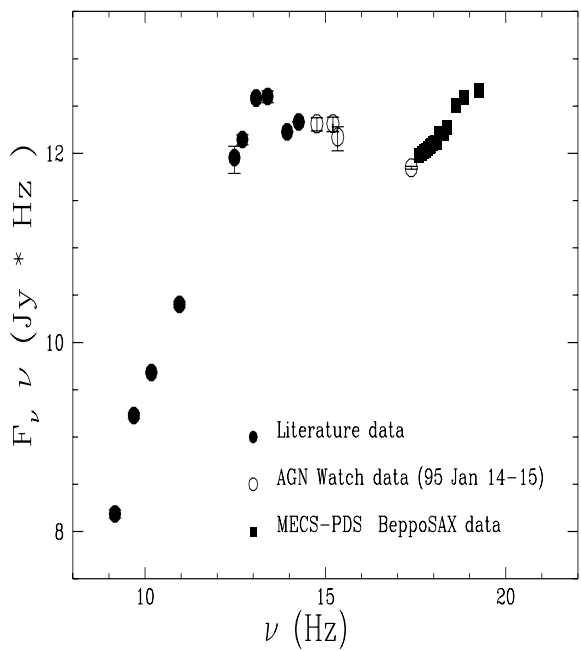,height=9.0cm,width=8.0cm, angle=-90}
\caption{Spectral energy distribution of 3C~390.3, 
from non-simultaneous observations. Radio-mm-infrared data
(filled circles) are taken from literature (see text).
Simultaneous optical UV and X-ray data (open circles) 
refer to the observation on 1995 January 14-15. 
The MECS and PDS data (filled squares) have been rebinned.}
\end{figure}
\noindent

Figure 6 shows that the energetics are dominated by the high energy end of the
power--law, and by the large IR emission. The power emitted in the UV is
definitely smaller than that in the X-ray to hard X-ray component. This is 
consistent with the results of Wo\'zniak et al. (1998), which show
that during spectral variations the total energy output in X--rays does not
change (see their Figure 5). 

All these facts indicate that the UV--emitting, optically thick gas must
subtend a relatively small solid angle to the X--ray source, otherwise strong
reprocessing would give rise to a thermal UV component that is energetically important
and efficient cooling would steepen the X--ray power--law. On the other hand,
as discussed in \S 3, the observed reflection hump and iron line need a fairly
 large covering factor of reprocessing material in order to be
accounted for. How can
these two, apparently contradictory, constraints be matched? 

There are basically two models under discussion to account for X-ray emission
from accretion disks in AGN: 

\begin{enumerate}

\item{A Seyfert--like model that assumes a geometrically thin accretion disk
(Shakura \& Sunyaev 1973), which is responsible for the UV thermal emission and
for the reprocessing of (part of) the X--ray photons. The high energy photons
are produced by an active corona embedding the inner portion of the cold
accretion disk (Liang 1979, Haardt $\&$ Maraschi 1991, 1993). In this class of
models, the Compton parameter is kept fixed by the energetic feedback linking
the disk and the corona. UV photons in the disk are produced by thermalization
of the absorbed X--rays, and X--rays in the corona are produced by inverse
Comptonization of the UV disk radiation;} 

\item{A hot accretion flow model, such as the original two--temperature solution
introduced by SLE or the ion--supported torus proposed by Ichimaru (1977), and
Rees et al. (1982), or its modern version, the ADAF (see NMQ for a review and
all the relevant references). In the ion--supported torus and in the ADAF,
relevant for low to modest accretion rates, the small gas density makes Coulomb
collisions very ineffective in transferring energy from the ions (which are
supposed to be directly energized by viscous stresses) to the electrons which
bear the ultimate responsibility of radiating away the heat and cooling the
gas. The direct consequence is the formation of a hot, two--temperature plasma
in the inner region of the flow. In contrast, in the SLE solution the
energy deposited in the gas is assumed to be locally radiated. In both classes
of models, at larger distances from the black hole, the flow is thought to be
described by a standard cooling--dominated thin disk (e.g., Mahadevan 1977).}

\end{enumerate} 

From the point of view of the formation of the radiation spectrum, the main 
difference between the two pictures is the presence (in the disk--corona 
system) or the absence (in the ion--supported torus) of optically thick cold 
matter close to the X--ray source providing (or not) soft photons 
for the Comptonization mechanism. In the absence of a soft photon input
from thermal optically thick gas, the seed photons are provided by 
cyclo-synchrotron radiation by the electrons themselves, yielding a power law 
which extends from the far IR up to hard X-rays with spectral index fixed by the
accretion rate. 

In the case of 3C~390.3, although a disk--corona model could explain the strong
correlation between the IUE and X--ray light curves, the absence of a soft
excess and the weakness of a possible blue bump argues against a large
fraction of reprocessed radiation. An optically thick corona radiating all the
available gravitational power could in principle scatter off all the black body
photons from the accretion disk and hence produce a unique Compton--scattered
power law (Haardt \& Maraschi, 1993), with weak or absent signature of thermal
emission. In this case, however, such a disk--corona system would give rise to
a power law spectrum steeper than observed ($\Gamma\gta 2$). In order to
produce an X--ray power--law as flat as observed, one has to assume a 
photon--starved corona. Thus the required geometry is one in which the UV--emitting
layer is at least partly external with respect to the region containing the hot
electrons. 

A completely hot inner flow, on the other hand, is a plausible description of
the nuclear region. A hot inner region can in fact explain the lack of soft
excess and the weak UV bump, which might still be the signature of an external
standard cold thin disk. The ion--supported torus, or the ADAF, is one of the
possible stable configurations of gas accreting onto a black hole. The optical
to X--ray radiation is due to Compton cooling of the hot thermal electrons
(with temperature $\sim 10^9$~K), scattering off soft free--free and
cyclo--synchrotron photons. If the accretion rate is high (but still below the
ADAF critical accretion rate, $\dot m_{\rm crit}\simeq 0.1 \dot m_{\rm Edd}$,
see NMQ), the bremsstrahlung contribution to the X--ray spectrum is negligible,
and the X--ray continuum is hard ($\Gamma< 2$). In the case of 3C~390.3, the
bolometric luminosity estimated from Figure~6 is $L\sim 3\times10^{45}$ erg
sec$^{-1}$. If we assume a central mass of $\sim 1-4\times 10^{8}$ M$_{\odot}$
(Wamsteker et al. 1997), the luminosity in Eddington units is $L/L_{Edd}\sim
0.05-0.2$., which would place 3C~390.3 in the range of high accretion rate
ADAFs, consistent with the hard X--ray continuum. We note nevertheless that
with this model the similarity of the spectral shape of 3C~390.3 to that of
Seyfert galaxies would be coincidental. 

If the accretion flow at larger radii is in the form of a standard thin disk,
the weak blue bump could be also explained, as due to local energy release, and
reprocessing of the (small) fraction of the X--rays intercepted and reprocessed
by the cold matter. A flat infinite disk illuminated by a central hot
ion--supported torus can not intercept more than 25$\%$ of the primary
continuum (Chen \& Halpern 1989), adequate for the observed UV emission in
this case, but not for the observed reflection component and for the iron line
EW. It is then necessary that further cold material, encircling the central
source, is shaped like a warped disk or a thick dusty torus at parsec
distances. These geometries can ensure large covering factors, and produce a broad
reflection hump practically indistinguishable (at this sensitivity level)
from that arising from an infinite plane parallel medium (Ghisellini, Haardt \&
Matt, 1994; Krolik, Madau \& Zycki, 1994). In particular, the K$_{\alpha}$
feature is expected to be narrow, in agreement with the iron line profiles
observed by \sax~ ($\sigma=70^{-70}_{+207}$ eV) and {\it ASCA} (Eracleous et al.
1997). The energetics would not be a problem in this picture, as most of the
X--rays are absorbed at large distances from the source, and then re--emitted
as IR radiation, rather than in the form of a UV bump as in the case of a
Seyfert--like geometry. A simple test of this model would be the absence of
short term variability both in the intensity of the Fe Line (Wozniak et al. 1998)
and of the reflection component. For the latter, further \sax~ observations 
would be valuable.

An inner hot torus 
surrounded by an outer cold thin disk was already proposed by Chen and Halpern
(1989) to explain the optical properties of BLRGs with double peaked emission
lines and in particular of 3C~390.3. Our observations independently
strengthen this picture. Whether this configuration can be consistent with the SLE
solution, with the ion--supported torus, and/or with the ADAF accretion models 
is a matter of future investigations, beyond the scope of the present paper.

Finally, an important result of our studies is the discovery in 3C~390.3 of
temporal variations of the local column density. 
The origin of the \nh~ variability is unknown. The presence of a warm absorber,
usually invoked to explain modification of the column density in Seyfert
galaxies, seems unlikely in 3C~390.3. The lack of features in
absorption/emission in the soft X--ray spectrum and the absence of any
correlation between the \nh~ values and the X--ray flux argue against this
possibility. The long term variations can be better explained by 
geometrical modifications of a cold absorber.

\section{Summary} 

We have presented \sax~observation of the BLRG 3C~390.3. The X--ray data are
well represented by an absorbed hard power law, plus a narrow $K_{\alpha}$ iron
line and a Compton reflection component. No X-ray soft excess is required 
by the data in agreement with the results of previous X-ray satellites.

Considering also that a weak blue bump seems to characterize this source, 
we conclude that:

\begin{enumerate}
\item{Beamed non--thermal X--ray continuum does not significantly 
contaminate the nuclear emission.}
\item{Material able to absorb and thermalize the 
X--rays is not present close to the X--ray source (otherwise we would 
observe a luminous UV to soft X--ray thermal component).} 
\item{However, such material must lie at some distance and intercept 
a fraction $\sim$ 50\% of the X--rays (otherwise we would not observe 
any reflection hump and iron line).}
\end{enumerate}

The model that better accounts for these results is a hot, X--ray emitting
inner accretion flow. Roughly half of the X--ray radiation illuminates outer,
cold, IR emitting material, possibly in the form of a warped disk and/or a
dusty torus, producing the reflection hump and the narrow emission line. At
this stage, it is premature to assess the nature of the hot inner flow.
Different models (i.e., the ion--supported torus, the ADAF, or the SLE) could 
possibly be
compared by studying UV--to--X--ray spectral variability on long time
scales. For this aim, further broad--band observations are most welcome. 

Finally we have presented a study of the time variations of \nh~, based on all
the available historical data. We have shown, for the first time, that the
column density in this source changes on time scale of years independent of 
variations in
the X--ray flux. It is difficult to relate the variations of the absorber to
changes of the ionization state of the medium along our line of sight. 
Geometrical modifications of the absorber would better account for the 
long term \nh~variations.

\begin{acknowledgements}
We are grateful to G. Palumbo for critical
reading of the manuscript.
We would like to thank T. Yaqoob, W. Wamsteker for useful
discussions, and L. Angelini for help in reducing the archival EXOSAT data.
We wish to thank the referee for useful comments on the manuscript. 
\end{acknowledgements}

\end{document}